\documentclass[aps,twocolumn,floatfix,showpacs]{revtex4}
\usepackage{graphicx}

\begin{document}

\title{An analytically solvable model of probabilistic network dynamics}

\author{M.A.M. de Aguiar$^{1,2}$, Irving R. Epstein$^{1,3}$ and Yaneer Bar-Yam$^1$}

\affiliation{$^1$ New England Complex Systems Institute,
Cambridge, Massachusetts 02138\\ $^2$Instituto de F\'isica Gleb
Wataghin, Universidade Estadual de Campinas, 13083-970 Campinas,
S\~ao Paulo, Brazil\\ $^3$ Department of Chemistry, MS015,
Brandeis University, Waltham, MA 02454}

\begin{abstract}

We present a simple model of network dynamics that can be solved
analytically for uniform networks. We obtain the dynamics of
response of the system to perturbations. The analytical solution
is an excellent approximation for random networks. A comparison
with the scale-free network, though qualitatively similar, shows
the effect of distinct topology.

\end{abstract}

\pacs{87.10.+e,02.50.Ey,84.35.+i}

\maketitle

Recent advances in the understanding of complex social
\cite{wasserman94}, biological \cite{fw1}, and technological
\cite{faloutsos} systems have revealed widespread if not universal
properties of the topology of networks of association, interaction
and communication. These properties, include small-world global
connectivity \cite{watts98}, scale-free local connectivity
distribution \cite{bararev}, and characteristic local motif
structures \cite{milo}. Central to our understanding of complex
systems \cite{baryam97} is characterizing their response to
environmental stimuli. While much of the focus has been on
robustness to random perturbation or directed attack,
\cite{albert2000} the effectiveness of response requires
satisfying a wider range of requirements including, for example,
sensitivity to particular stimuli \cite{baryam2004}. Indeed, one
of the main functions of biological and social systems is the
detection of specific stimuli that require collective (large
scale) response in seeking desirable resources (foraging) or
responding to dangers (``fight or flight"). Thus understanding the
nature of system function and behavior from topological structure
requires mapping the interaction structure given by a topological
network onto the dynamics of system response \cite{zanette}.
Insofar as the network of interactions obtained in recent research
characterizes the internal interactions of a system, these
interactions must provide key information about the dynamics of
response to external perturbations. Therefore, the construction of
solvable dynamic models is essential for understanding the general
features of the problem. In this Letter we propose such a model of
probabilistic network dynamics, and we solve it analytically for
uniform networks.

We consider a general network with $N$ nodes. To each node $i$ is
assigned an {\it internal} state $\sigma_i$ that can take the
values $0$ or $1$. At each time step the state of a node is
updated according to the following rule: either the state does not
change, which happens with probability $p$ or, with probability
$(1-p)$, it copies the state of one of its neighbors.  This
process describes, for example, the behavior of a group of high
school students choosing to adopt one style of dress or another,
or the propagation of a mutation through a species \cite{rauch04}.
Since the states of a node are abstract labels, the change of one
node to adopt the state of another can be considered a general
model of influence propagation, with each node state a label for
its own relevant physical property.

The $2^N$ states of the network can be labeled by a string of
zeroes and ones describing the internal state of each node in
sequence $(\sigma_{N\!-\!1}\sigma_{N\!-\!2} \dots \sigma_1
\sigma_0)$. Alternatively, the states can be labeled by integers
via $x=\sum_{j=0}^{N-1} \sigma_j 2^j$, with $x$ varying between 0
and $2^N-1$.

Let $P_t(x)$ be the probability of finding the network in the
state $x$ at time $t$ and let the network evolve through
asynchronous updates, where a single node is allowed to change at
each time step. To find how this probability changes with time we
define the auxiliary state $\tilde{x}_k$ which is equal to $x$ at
all nodes except at node $k$, which has the opposite internal
state. The probability of finding the network in the state $x$ at
time $t+1$ can now be written as a sum of three terms: (a) the
probability that the network was in state $x$ at time $t$ and that
the selected node did not change plus (b) the probability that it
was in the state $x$ and the selected node copied the state of an
identical neighbor plus (c) the probability that the network was
in the state $\tilde{x}_k$ at time $t$ and that the node $k$ was
selected and its state $\tilde{\sigma}_k=1-\sigma_k$ changed to
$\sigma_k$:
\begin{displaymath}
\begin{array}{ll}
P_{t+1}(x) &=  p P_t(x) +\displaystyle{\frac{(1-p)}{N} \sum_k
\left\{ P_t(x) ~ \rm{Prob}[\sigma_k\rightarrow \sigma_k] \right.}  \\
 &+ \; \displaystyle{\left. P_t(\tilde{x}_k)~\rm{Prob}[\tilde{\sigma}_k\rightarrow
 \sigma_k] \right\}}\;.
\end{array}
\end{displaymath}

The probability $\rm{Prob}[\sigma_k\rightarrow \sigma_k]$ is just
the number of neighbors of node $k$ in the state $\sigma_k$
divided by the total number of neighbors (the degree) $d_k =
\sum_{i=0}^{N-1} C_{ik}$, where $C_{ik}$ is the connectivity (or
adjacency) matrix. This can be written as
\begin{displaymath}
\frac{1}{d_k} \sum_{i=0}^{N-1} C_{ik} |1-\sigma_i-\sigma_k|
\end{displaymath}
The probability $\rm{Prob}[\tilde{\sigma}_k\rightarrow \sigma_k]$
is also given by this formula, since $\tilde{\sigma}_k=1-\sigma_k$
and $\tilde{\sigma}_i=\sigma_i$ for $i\neq k$. Using these
relations, we obtain the following master equation for the network
dynamics:
\begin{equation}
\begin{array}{ll}
P_{t+1}(x) =&  p P_t(x) \; +  \displaystyle{\frac{(1-p)}{N}}
\sum_{k=0}^{N-1} \frac{1}{d_k} \sum_{i=0}^{N-1} C_{ik}  \times \\
&\displaystyle{ |1-\sigma_i-\sigma_k|  \left[ P_t(x) +
P_t(\tilde{x}_k)\right]} \;.
\end{array}
\label{masterg}
\end{equation}

Finding $P_t(x)$ for networks with arbitrary topologies can be
very difficult. However, the problem can be completely solved for
fully connected networks, where $d_k = N-1$. In this case the
nodes are indistinguishable from each other and the states of the
network can be labeled simply by counting the number of nodes in
the internal state $1$, given by $n(x)=\sum_i \sigma_i$. The
probability of finding the network in the state labeled by $n$ is
related to $P(x)$ by
\begin{equation}
P(n(x)) = P(x) \, B(N,n)  \label{mult}
\end{equation}
where $B(N,n)=N!/[n!(N-n)!]$ is a binomial coefficient. We now
simplify the last two terms on the right of Eq.(\ref{masterg}). To
do this we separate the sum over $k$ into the cases $\sigma_k=1$
and $\sigma_k=0$. For the first of these terms we obtain
\begin{displaymath}
\begin{array}{l}
 \frac{1-p}{N(N-1)} \sum_k \sum_i [C_{ik} \sigma_i+C_{ik} (1-\sigma_i)] P_t(x)   \\
=\frac{1-p}{N(N-1)}\left[n(n-1) + (N-n)(N-n-1) \right] P_t(x) \;.
\end{array}
\end{displaymath}

For the third term we observe that $P_t(\tilde{x}_k)$ corresponds
to the state $n-1$ if $\sigma_k=1$ and to $n+1$ if $\sigma_k=0$.
When we separate the sum over $k$ into the cases $\sigma_k=1$ and
$\sigma_k=0$ we write $P_t(\tilde{x}_k;1)$ and
$P_t(\tilde{x}_k;0)$ respectively. We obtain
\begin{displaymath}
\frac{(1-p)}{N(N-1)}[n(n-1) P_t(\tilde{x}_k;1) + (N-n)(N-n-1)
P_t(\tilde{x}_k;0)]\;.
\end{displaymath}

Substituting these terms into the master equation and multiplying
both sides by $B(N,n)$ we obtain, after some simplification,
\begin{equation}
\begin{array}{ll}
P_{t+1}(n) &= p P_t(n) + \frac{1-p}{N(N-1)} \times \\
&\left\{ [n(n-1)+(N-n)(N-n-1)]P_t (n) \right.  \\
 &+ (N-n+1)(n-1) P_t(n-1)\\
&\left. +(N-n-1)(n+1)P_t(n+1) \right\} \;. \label{masterf}
\end{array}
\end{equation}

For {\it uniform networks} where $d_k=d_0$ is the same for all
nodes, if $d_0 < N-1$ states with the same $n(x)$ can be
distinguished  by the way the internal states with $\sigma_k=1$
are distributed among those with $\sigma_k=0$. However, if we
combine states with the same $n(x)$, the procedure described above
can still be applied. In this case the factor $d_k$ in the
denominator is replaced by $d_0$. However, on the average (with
respect to the different states labeled by $n$), the counting of
the number of neighbors must be multiplied by $d_0/(N-1)$, so that
$d_0$ cancels and we get $(N-1)$ back in the denominator.
Therefore equation (\ref{masterf}) holds in this case as well. For
random networks the degree of each node is nearly constant, and we
can still use equation (\ref{masterf}) as an approximation for the
dynamics.

We now proceed to the calculation of the transition probabilities.
The probabilities $P_t(n)$ define a vector $P_t$ of $N+1$
components. The master equation (\ref{masterf}) can be written in
matrix form as $P_{t+1} = U P_t$ where the {\it evolution matrix}
$U$ is tridiagonal. The propagation of an initial probability
vector requires the calculation of powers of $U$. Alternatively,
we can diagonalize $U$ and use its eigenvectors as a basis. This
approach has been used \cite{cannings} to calculate the
eigenvalues of the transition matrix for certain population
models. Here we shall calculate not only the eigenvalues but also
the eigenvectors, obtaining the complete solution of the dynamical
problem.

The eigenvalues of $U$ can be calculated for small matrices and
extrapolated to matrices of arbitrary size. They are given by
\begin{displaymath}
\lambda_r = 1-\frac{1-p}{N(N-1)}r(r-1) \;.
\end{displaymath}
with $r=0,1,\dots,N$. The only degeneracy occurs for
$\lambda_{0}=\lambda_1=1$. The other eigenvalues are all smaller
than $1$ and decrease towards $\lambda_N = p$.

Since $U$ is not symmetric, its eigenvectors do not form an
orthogonal set. Let $|a_r\rangle$ and $\langle b_r|$ be the right
and left eigenvectors of $U$, with components $a_{rm}$ and
$b_{rm}$. Then
\begin{displaymath}
\sum_{r=0}^N \frac{1}{\Gamma_r}|a_r\rangle \langle b_r| = 1 \;.
\end{displaymath}
where $=\langle b_{r'}|a_r\rangle=\Gamma_r \delta_{rr'}$ and
$\Gamma_r = \sum_m a_{rm} b_{rm}$.

An initial vector $|v(0)\rangle$ containing the information about
the probability of the different states at time zero can be
projected using this resolution of unity and easily evolved:
\begin{displaymath}
|v(t)\rangle = U^t |v(0)\rangle = \sum_{r=0}^N \frac{1}{\Gamma_r}
\langle b_r |v(0)\rangle \, \lambda_r^t \, |a_r\rangle \;.
\end{displaymath}

The transition probability between two network states with $n=M$
and $n=L$ after a time $t$ can now be calculated by taking the
components of the initial vector as $v_m(0) = \delta_{M,m}$ and
projecting the evolved state onto the state with components
$\delta_{L,m}$:
\begin{displaymath}
P(L,t;M,0) = \sum_{r=0}^N \frac{1}{\Gamma_r} b_{rM} a_{rL}
\lambda_r^t \;.
\end{displaymath}

The coefficients $a_{rm}$ follow a recursion relation that can be
derived directly from the eigenvalue equation for $U$. For $r=0$
and $r=1$ the eigenvectors can be found immediately:
\begin{equation}
|a_0\rangle = \left(
\begin{array}{c} 1 \\ 0 \\ 0 \\ \vdots \\ 0 \\ 1 \end{array}
\right), \quad |a_1\rangle = \left(
\begin{array}{c} 1 \\ 0 \\ 0 \\ \vdots \\ 0 \\ -1 \end{array}
\right) \label{rvec01}
\end{equation}
and
\begin{equation}
\begin{array}{l}
|b_0\rangle = \left( 1 \; 1 \; 1 \; \dots \; 1 \; 1 \right), \\ \\
|b_1\rangle = \left(N \;\;\; N\!-\!2 \;\;\; N\!-\!4 \;\;\; \dots
\;\; -\!N\!+\!2 \;\;\; -\!N \right) \;.
\end{array} \label{lvec01}
\end{equation}

In order to calculate the remaining eigenvectors, we define the
auxiliary eigenvalues $\mu_r$ by
\begin{equation}
\mu_r = (1-\lambda_r)\frac{N(N-1)}{1-p}= r(r-1)
\end{equation}
and the auxiliary coefficients
\begin{equation}
A_{rm} = m(N-m) a_{rm} \qquad m=1,2,\dots,N-1 \;.
\end{equation}
The recursion relation for the $A_{rm}$ can be written explicitly
as:
\begin{equation}
\label{eq3} A_{r\,m+1} -2 A_{rm}+A_{r\,m-1} =
-\frac{\mu_r}{N}\left( \frac{A_{rm}}{m} +
\frac{A_{rm}}{N\!-\!m}\right) \;.
\end{equation}

A generating function is now defined as
\begin{displaymath}
f_r(x) = \sum_{m=1}^{N-1} A_{rm} x^m
\end{displaymath}
(note that $A_{r0}=A_{rN}=0$). Multiplying eq.(\ref{eq3}) by $x^m$
and summing over $m$ we get, on the left side,
\begin{displaymath}
\frac{f_r}{x}(1-x)^2 -A_{r1}+A_{r\,N\!-\!1} x^N\;.
\end{displaymath}
In order to write down the right side of eq.(\ref{eq3}) we define
the auxiliary functions
\begin{equation}
g_r(x) = \sum_{m=1}^{N-1} \frac{A_{rm}}{m} x^m \quad , \quad
h_r(x) = \sum_{m=1}^{N-1} \frac{A_{rm}}{N-m} x^m .
\end{equation}
It is easy to check that $d g_r/d x = f_r/x$ and $d h_r/d x = N
h_r/x - f_r/x$. After multiplying Eq.(\ref{eq3}) by $x^m$ and
summing over $m$, we differentiate both sides with respect to $x$
to obtain
\begin{equation}
\frac{d}{dx}\left[\frac{(1-x)^2}{x}f_r(x)-A_{r\,N\!-\!1}x^N\right]
= -\frac{\mu_r}{x} h_r(x) \label{eqf} .
\end{equation}

The solution of the differential equation for $h_r$ can be
obtained in terms of its Green function, satisfying $d G/d x - N
G/x = \delta(x-y)$. In this case $G$ is given by $(x/y)^N$ if
$x>y$ and zero otherwise. Therefore,
\begin{equation}
h_r(x) = x^N\left(\alpha -\int_{-\infty}^x \frac{f_r(y)}{y^{N+1}}
~dy \right) \;. \label{eqhf}
\end{equation}

Substituting Eq.(\ref{eqhf}) into (\ref{eqf}), re-arranging the
terms and differentiating once again with respect to $x$, we
obtain
\begin{displaymath}
\frac{d}{dx}\left[
\frac{1}{x^{N-1}}\frac{d}{dx}\left(\frac{(1-x)^2}{x}f_r(x)\right)\right]
= \mu_r  \frac{f_r(x)}{x^{N+1}} \;. \label{eqf3}
\end{displaymath}
Finally, defining
\begin{equation}
F_r(x) = \frac{(1-x)^2}{x}f_r(x) \label{eqff}
\end{equation}
we obtain the differential equation
\begin{equation}
F_r'' - \frac{N-1}{x} F_r' - \frac{\mu_r}{x} \frac{F_r}{(1-x)^2} =
0 \;. \label{eqdif}
\end{equation}

Letting $\phi_r(x) = \sum a_{rm} x^m$ then
\begin{displaymath}
\phi_r(x) = \sum \frac{A_{rm}}{m(N-m)} x^m
     =\frac{1}{N}(g_r + h_r) \;.
\end{displaymath}
Differentiating with respect to $x$, using Eq.(\ref{eqhf}),
dividing by $x^{N-1}$ and differentiating again, we find
\begin{displaymath}
\phi_r'' - \frac{N-1}{x} \phi_r' - \frac{1}{x} \frac{F_r}{(1-x)^2}
= 0 \;. \label{eqdifg}
\end{displaymath}
Comparing with Eq.(\ref{eqdif}) we see that $\phi_r = -
F_r/\mu_r$. Therefore, except for a normalization, {\it the
generating function for the coefficients $a_{rm}$, $\phi_r(x)$, is
equal to $F_r(x)$.}

For $r=0$ or $r=1$, $\mu_r=0$ and the two independent solutions of
eq.(\ref{eqdif}) are $F_0(x) = 1 + x^N$ and $F_1(x) = 1 - x^N$,
which correspond to the two degenerate eigenvectors $|a_0\rangle$
and $|a_1\rangle$. For $r=2$ and $r=3$ the solution can also be
found explicitly; the general formula can then be extrapolated
from these simple cases. We find
\begin{equation}
F_{r}(x) = (1-x)^{1-r} \left[1+ \sum_{p=1}^{r-1} d_{rp} \, x^p
\right] \;.
\end{equation}
with
\begin{displaymath}
d_{rp} = (-1)^p \frac{B(r-1,p)~B(N+r-1,p)}{B(N-1,p)} \;.
\end{displaymath}

Finally, the coefficients of the $r$-th eigenvector are given by
\begin{displaymath}
a_{rm} = \frac{1}{m!} \left. \frac{d^m F_r(x)}{d x^m}\right|_{x=0}
\;.
\end{displaymath}
Since $F_r$ is the product of two simple functions, its derivative
can be calculated explicitly at $x=0$. Writing $F_r(x) = N_r(x)
Q_r(x)$ with
\begin{displaymath}
Q_r(x) = (1-x)^{1-r} \quad \mbox{and} \quad N_r(x) = 1 + \sum
d_{rp} \, x^p
\end{displaymath}
we find
\begin{displaymath}
a_{rm} = \frac{1}{m!} \sum_{p=0}^m B(m,p) \left.
\frac{d^{m-p}N_r}{dx^{m-p}}\right|_{(x=0)}\left. \frac{d^p
Q_r}{dx^p}\right|_{(x=0)} \;.
\end{displaymath}
Working out the derivatives we find the explicit formula valid for
$r \geq 2$.
\begin{equation}
a_{rm} = \displaystyle{\sum_{p=0}^{r-1} B(m-p+r-2,r-2) \, d_{rp}}
\label{coefa}
\end{equation}
for $m=1,2,\dots,N-1$, with $a_{r0}=1$ and $a_{rN}=(-1)^{r}$.

From the recursion relations we find that the coefficients of the
left eigenvectors are given by
\begin{equation}
b_{rm} = a_{rm} \, [m(N-m)/N] \label{coefb}
\end{equation}
for $m=1,2,\dots,N-1$, with $b_{r0}=b_{rN}=0$. Finally, the
normalization factors $\Gamma_r$ can also be obtained explicitly:
\begin{equation}
\Gamma_r= \frac{r!~ B(N+r-1,r)}{(2r-1)~ B(N,r)} \label{gammar}
\end{equation}
for $r=2,\dots,N$ and $\Gamma_0=2$, $\Gamma_1=2N$.

We are finally in a position to state some important results
concerning the dynamics. The transition probability of starting
the network at time zero with $n=M$ and finding it at a later time
$t$ with $n=L$ can be computed using Eqs.(\ref{rvec01}),
(\ref{lvec01}), (\ref{coefa}), (\ref{coefb}) and (\ref{gammar}):
\begin{displaymath}
\begin{array}{l}
P(L,t;M,0) = \left[\frac{N-M}{N} - \frac{3M(N-M)}{(N+1)N}
\lambda_2^t \right]\delta_{L0} \, + \\  \left[\frac{M}{N}-
\frac{3M(N-M)}{(N+1)N} \lambda_2^t \right]\delta_{LN} \, + \\
\frac{6M(N-M)}{N(N^2-1)}(1\!-\!\delta_{L0})(1\!-\!\delta_{LN})
\lambda_2^t +\, \sum_{r=3}^{N} \frac{1}{\Gamma_r} b_{rM} a_{rL}
\lambda_r^t .
\end{array}
\end{displaymath}

Since all eigenvalues (except for $\lambda_0$ and $\lambda_1$) are
smaller than one, in the limit of long times the transition
probability is dominated by $\lambda_0=\lambda_1=1$ and by the
largest non-trivial eigenvalue, $\lambda_2$, whose contributions
we have written down explicitly. For large networks we can
approximate $\lambda_2^t \approx \exp{-[2(1-p)t/N^2]}$, so that
the characteristic duration of the transition process is $\tau =
N^2/(1-p)$, which increases with the square of the network size.
\begin{figure}
   \includegraphics[width=6.cm,angle=-90]{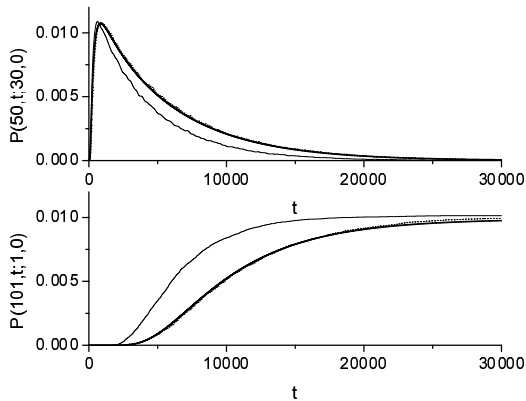}
  \caption{Transition probabilities for a network with $N=101$ and $p=0.1$.
  The lines correspond to our theoretical calculation (thick) and to
  simulations for random (dotted) and scale-free (thin) networks, both with
  an average of $6$ connections per node. The numerical probabilities were
  computed running the simulations $2\times 10^5$ times. The dotted line is
  nearly indistinguishable from the thick line.}
  \label{fig1}
\end{figure}

The only two possible asymptotic states are $n=0$ and $n=N$, whose
transition probabilities from an initial state $n=M$ are $(N-M)/N$
and $M/N$ respectively. A typical transition probability
$P(L,t;M,0)$ for $L,M\neq 0,N$ starts at zero if $L\neq M$,
reaches a maximum and decreases back to zero. This represents the
probability that a perturbation initially affecting M nodes will
lead to a response by L nodes at a time t later. Figure 1 shows an
example for a network with $N=101$ nodes and $p=0.1$. The estimate
$\tau\approx 11000$ works well for all the transition
probabilities shown. These results show that the theoretical model
for fully connected networks is an excellent approximation for the
average behavior of sparsely connected random networks. The theory
also reproduces qualitatively the behavior of scale-free networks.
The deviations from the theory in this case reflect the
significant topological differences between the two networks. As a
final remark we note that the case of more internal states per
node can in principle be treated in a similar fashion. However,
although Eq.(\ref{masterg}) can be easily generalized,
Eq.(\ref{masterf}) would be more complicated, since it would take
more than a single integer $n(x)$ to describe the network state.

M.A.M.A. acknowledges financial support from CNPq and FAPESP.
I.R.E. was supported in part by the David and Lucile Packard
Foundation.

\end{document}